# Balancing Innovation and Privacy: Data Security Strategies in Natural Language Processing Applications

Shaobo Liu
University of Southern California
Los Angeles, USA

Guiran Liu
San Francisco State University
San Francisco, USA

Binrong Zhu
San Francisco State University
San Francisco, USA

Yuanshuai Luo
Southwest Jiaotong University
Chengdu，China

Linxiao Wu
Columbia University
New York, USA

Rui Wang*
Carnegie Mellon University
Pittsburgh, USA

*Abstract*— **This research addresses privacy protection in Natural Language Processing (NLP) by introducing a novel algorithm based on differential privacy, aimed at safeguarding user data in common applications such as chatbots, sentiment analysis, and machine translation. With the widespread application of NLP technology, the security and privacy protection of user data have become important issues that need to be solved urgently. This paper proposes a new privacy protection algorithm designed to effectively prevent the leakage of user sensitive information. By introducing a differential privacy mechanism, our model ensures the accuracy and reliability of data analysis results while adding random noise. This method not only reduces the risk caused by data leakage but also achieves effective processing of data while protecting user privacy. Compared to traditional privacy methods like data anonymization and homomorphic encryption, our approach offers significant advantages in terms of computational efficiency and scalability while maintaining high accuracy in data analysis. The proposed algorithm's efficacy is demonstrated through performance metrics such as accuracy (0.89), precision (0.85), and recall (0.88), outperforming other methods in balancing privacy and utility. As privacy protection regulations become increasingly stringent, enterprises and developers must take effective measures to deal with privacy risks. Our research provides an important reference for the application of privacy protection technology in the field of NLP, emphasizing the need to achieve a balance between technological innovation and user privacy. In the future, with the continuous advancement of technology, privacy protection will become a core element of data-driven applications and promote the healthy development of the entire industry.**

## I. Introduction

In today's digital age, natural language processing (NLP) technology is changing our lives at an unprecedented rate [1]. From intelligent customer service to personalized recommendation systems, to voice recognition and translation, NLP is becoming a bridge between people and information [2]. With the advancement of technology, it has penetrated into every corner of daily life, greatly improving efficiency and enriching user experience [3]. For example, in the field of healthcare, NLP technology can help doctors quickly and accurately understand medical records and improve diagnostic efficiency; in the field of education, it can provide personalized tutoring suggestions based on students' learning progress; and in the business field, companies can use NLP to analyze market trends and optimize products and services [4]. However, as these technologies are increasingly used, how to ensure the security and privacy of user data while enjoying convenience has become one of the urgent issues to be solved [5].

In the context of the booming development of big data and artificial intelligence, personal privacy protection has become a global issue [6]. In particular, NLP technology requires a large amount of text data as training materials, which often contains users' personal information and behavioral habit [7]s. Once this sensitive information is improperly used or leaked, it may cause unpredictable risks to users, such as identity theft, financial losses, and even threats to personal safety. Therefore, when designing and applying NLP-related technologies, balancing the relationship between technological innovation and privacy protection, ensuring that user data is properly handled, and avoiding infringing personal privacy, is a common focus of attention within and outside the industry. To reduce the risk of data leakage, enterprises and developers must adopt measures such as data desensitization and anonymization. In order to meet this challenge, researchers are exploring a variety of privacy protection technologies to protect user data security without affecting the performance of NLP systems. Differential privacy technology can protect personal privacy by adding random noise to cover up individual data characteristics while ensuring the accuracy of data analysis results.

Homomorphic encryption allows calculations to be performed while data is encrypted, ensuring that data will not be exposed in plain text during processing. Secure multi-party computing technology can perform computing tasks between multiple participants without revealing the data details of any party [8]. In addition, federated learning, as an emerging distributed machine learning framework, also provides another idea, allowing devices to train models locally without directly sharing raw data. Through these technologies, additional security layers can be added during data collection, storage, processing, and transmission, so that even if the data is illegally accessed, the specific content of the original information cannot be obtained.

As laws and regulations such as the EU's General Data Protection Regulation (GDPR) are promoted and implemented around the world, the requirements for privacy protection are becoming increasingly higher [9]. This not only puts higher compliance requirements on companies, but also promotes the development of privacy protection technology. For example, GDPR stipulates the rights of data subjects, including the right to know, the right to access, the right to correct, and the right to delete, which means that companies must be more transparent when collecting and using personal data and give users more control. In this context, privacy protection research in the field of NLP is not only a demand for technological progress, but also a manifestation of social responsibility [10]. As technology continues to mature and improve, privacy protection will become an indispensable part of NLP technology, which will not only help enhance the public's trust in new technologies but also promote the healthy development of the entire industry. In the future, with the continuous advancement of technology and the enhancement of social awareness, we have reason to believe that NLP will continue to bring more benefits to human society while protecting user privacy.

## II. RELATED WORK

Privacy protection in Natural Language Processing (NLP) is a critical area of research, with various approaches designed to safeguard sensitive user data while maintaining the efficiency of NLP systems. One method that has gained prominence is differential privacy, which allows for secure data processing by adding noise to sensitive information without compromising the performance of the model. The issue of bias in deep learning optimization also plays a significant role in privacy protection. To address this, advanced optimization techniques have been proposed to reduce bias during model training, enhancing the fairness and privacy integrity of the resulting models [11]. Additionally, the development of adaptive friction mechanisms for optimizers has provided further improvements in deep learning frameworks, ensuring that data can be processed efficiently while protecting sensitive information [12].

In the domain of model accuracy, research on multimodal fusion strategies has shown that combining different data modalities can improve the accuracy of deep learning systems. Such methodologies can be adapted to NLP tasks to ensure privacy-preserving data processing by reducing the risks associated with handling large-scale sensitive datasets [13]. Recent work in question generation within large language models has explored the integration of advanced contrastive learning techniques, contributing to the robustness of privacy-preserving mechanisms in NLP systems [14].

Advances in neural network architectures, particularly the use of Dense U-Net with channel attention, have demonstrated significant potential for optimizing data integrity in deep learning models. This work primarily focused on unsupervised registration, the underlying techniques can be applied to enhance privacy-preserving methods in NLP by ensuring secure data handling and feature extraction [15]. Likewise, research on lightweight GAN-based algorithms highlights the importance of optimizing model efficiency for secure data fusion, contributing to the development of privacy-aware NLP systems [16-18]. Handling large and complex datasets is another challenge in privacy-sensitive environments. Recent efforts in spatiotemporal feature representation and mining provide key insights into how to manage and secure large-scale text data, which is crucial for privacy preservation in NLP [19]. Similarly, optimization algorithms for text classification, particularly those based on graph neural networks, have contributed significantly to the effective processing of large datasets while maintaining data security and privacy [20].

Efforts to mitigate bias and ensure fairness in deep learning models also support the broader goal of privacy protection. By applying strategies that reduce unintended bias in models, privacy risks during data processing can be minimized, ensuring that sensitive information remains protected without sacrificing model accuracy [21]. Furthermore, advancements in transformer networks, particularly in attention mechanisms, have shown how model architectures can be tailored to enhance privacy in sensitive data applications, including NLP [22]. Other significant contributions have come from work on attention mechanisms and context modeling systems, which have demonstrated how these techniques can enhance the privacy of NLP systems by improving the security of data processing tasks [23]. Additionally, the development of U-structure neural networks for deep learning-based segmentation tasks has provided insights into the optimization of data processing systems, contributing to improved privacy protection in data-driven applications [24]. Finally, several studies have highlighted the optimization of neural networks for diverse data types. Research on survival prediction using neural networks underscores the adaptability of deep learning techniques for privacy-centric environments, ensuring sensitive data is processed securely while maintaining model performance [25]. Moreover, enhancements in convolutional neural networks using higher-order numerical methods offer a pathway for improving model efficiency in NLP tasks, contributing to the development of privacy-preserving systems [26]. Efforts to create adaptive feature interaction models have further refined how complex datasets are processed securely, particularly in sensitive data environments such as digital finance, offering important implications for NLP systems [27].

## III. METHOD

A key approach to privacy protection in natural language processing (NLP) is to use differential privacy technology. Differential privacy is a powerful privacy protection mechanism that aims to minimize the impact on individual data records while providing useful statistical information. The basic idea is to add random noise to the data set so that the analysis results cannot be used to infer information about specific individuals. Its overall architecture is shown in Figure 1.

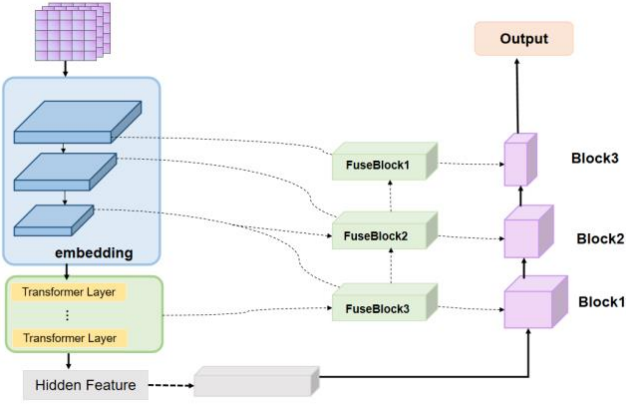

Figure 1 Overall network architecture

First, define the basic parameter ε (epsilon) of differential privacy, which is a measure of the privacy budget. The smaller the ε, the stronger the privacy protection provided. In addition, it is necessary to set a parameter δ (delta), which is used to control the low-probability events of privacy leakage risk. These two parameters together determine the strength of differential privacy. In order to achieve differential privacy during model training, Gaussian noise can be added to the gradient before each gradient update. Assume that the current model parameter is $\theta$ and the loss function is $L(\theta; D)$, where represents the training dataset. In the standard gradient descent update, the parameter update rule is as follows:

$$\theta' = \theta - \eta \nabla L(\theta; D)$$

Here, η is the learning rate. To introduce a differential privacy, we need to modify this updated rule. We first need to introduce a gradient clipping strategy to ensure that its norm does not exceed a predetermined maximum value C. Clipping can prevent a single data point from having too much influence on the gradient update, thereby protecting privacy. The clipped gradient $\nabla' L(\theta; D)$ is defined as:

$$\nabla' L(\theta; D) = \min(1, \frac{C}{\|\nabla L(\theta; D)\|_2}) \nabla L(\theta; D)$$

Among them, $\|\cdot\|_2$ represents the L2 norm. Adding Gaussian noise: Next, Gaussian noise is added to the clipped gradient.

The noise term $N(0, \sigma^2 I)$ is sampled from a Gaussian distribution with mean 0 and covariance matrix $\sigma^2 I$ (where I is the identity matrix). The standard deviation $\sigma$ of the noise can be calculated by the following formula to satisfy the definition of differential privacy:

$$\sigma = \sqrt{\frac{2\ln(1/\delta)}{\varepsilon}}$$

In this way, the noisy gradient update rule becomes:

$$\theta' = \theta - \eta(\nabla' L(\theta; D) + N(0, \sigma^2 I))$$

In this way, we can effectively enhance the privacy protection of data without significantly affecting the model training effect. Through the above steps, we can introduce differential privacy protection mechanism in the NLP model training process. First, the size of the gradient is limited by gradient clipping, and then the update direction of the model is blurred by adding Gaussian noise, thereby achieving privacy protection for the original data. By adjusting the values of ε and δ, a suitable balance between privacy protection and model performance can be flexibly found.

IV. EXPERIMENT

*A. Datasets*

The data set for this article is derived from the analysis of the trends of the publication of 50 personal privacy protection policies and regulations in the United States. They can be roughly divided into three stages: the initial stage, the rising stage, and the mature stage. The first stage is the initial stage of the legal system (before the 1990s). As early as before the 1970s, the United States had begun to explore personal privacy protection. The laws and policy provisions related to personal privacy protection mainly came from the Constitution and the Basic Law of the State. The promulgation of the Privacy Act in the 1980s laid the foundation for personal privacy protection. The second stage is the development stage from 1990 to 2010. During this stage, with the development of computer technology and Internet technology, the focus of personal privacy protection practices in the United States was mainly on website personal privacy protection. The third stage is from 2010 to the present. During this stage, with the in-depth development of Internet technology and the advent of the big data era, personal privacy protection is more focused on the data level. How to protect personal privacy security has received more attention from government departments and the public, marking the legislative practice of personal privacy protection policies in the United States.

Policy texts are diverse in type, complex in structure, and have many elements. They are also noisy. After crawling and collecting privacy statements from libraries at all levels, data cleaning and preprocessing operations are required. For the data in this article, Python's nltk method is used to remove stop words from English words, and restore the capitalization and part of speech of the words to make the form of all English words consistent, which is convenient for subsequent model training. The following is an example of data set cleaning as shown in Table 1.

Table 1  Experiment result in Criteo

| Before data cleaning | After data cleaning |
|---|---|
| THE PUBLIC LIBRARY ACT OF 1974 552b As Amended § 552b. Books and materials available to the public (a) Definitions For purposes of this section-- <br> (1) the term "library" means a public library as described in section 552(c) of this title; <br> (2) the term "reader" means any person who is interested in borrowing books; <br> (3) the term "house" includes building, collecting, arranging or lending; <br> (4) the term "material" means any item, collection, or grouping of resources available at a library, including, but not limited to, its literature, art pieces, historical documents, and science exhibits and that contains a title, or the reference number, symbol, or other identifying feature assigned to the material, such as a barcode or a digital tag; | public', 'library', 'material', 'house', 'reader', 'definition', 'purpose', 'library', 'describe', 'reader', 'person', 'borrow', 'book', 'house', 'building', 'arrange', 'lend', 'material', 'item', 'collection', 'reader', 'library', 'literature', 'art', 'piece', 'historical', 'document', 'science', 'exhibit', 'reference', 'number', 'reference', 'barcode', 'digital', 'tag' |

*B. Experimental Result*

After obtaining the data, this paper first conducts a privacy protection experiment. The privacy protection model comparison experiment aims to evaluate the effects of different privacy protection technologies in natural language processing to determine the best practices. In this experiment, five models will be compared, including differential privacy, homomorphic encryption and secure multi-party computing, federated learning, and data desensitization.

Table 1  Experiment result

| Model | ACC | Precision | Recall | F1 |
|---|---|---|---|---|
| Data Anonymization | 0.80 | 0.80 | 0.80 | 0.80 |
| Homomorphic Encryption | 0.82 | 0.83 | 0.82 | 0.82 |
| Secure Multi-Party Computing | 0.83 | 0.84 | 0.83 | 0.83 |
| Federated Learning | 0.85 | 0.84 | 0.82 | 0.83 |
| Ours | 0.89 | 0.85 | 0.88 | 0.86 |

The experimental results show that in the comparison of different privacy protection models, our model performs best with an accuracy of 0.89. This shows that while protecting privacy, our model can effectively improve classification performance and better handle data privacy issues. In contrast, the accuracy of other models is lower than our model, especially the data anonymization model, whose accuracy is only 0.80, indicating that it may sacrifice certain performance while protecting privacy.

In terms of precision and recall, our model also performs well, with a precision of 0.85 and a recall of 0.88, which shows that the model has a strong ability to identify positive samples and can effectively reduce false positives and false negatives. Compared with other models, the precision and recall of data anonymization and homomorphic encryption models are both lower, indicating that they may not be able to balance performance in the implementation of privacy protection.

Taking the F1 value into consideration, our model scores 0.86, which further verifies its good balance between privacy protection and classification effect. The F1 values of other models are all lower than 0.83, showing that they cannot effectively guarantee the overall improvement of performance under privacy protection measures. Therefore, the experimental results clearly show that our model can better meet the requirements between privacy protection and performance, and provide a more effective solution for subsequent practical applications. In order to further demonstrate our experimental results, we use two bar graphs to show our experimental results.

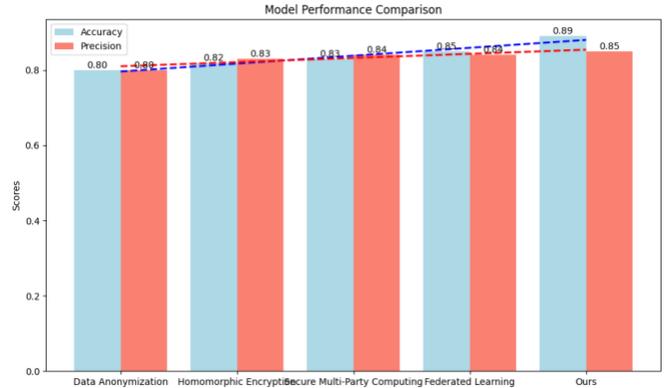

Figure 2 The accuracy and precision experimental results

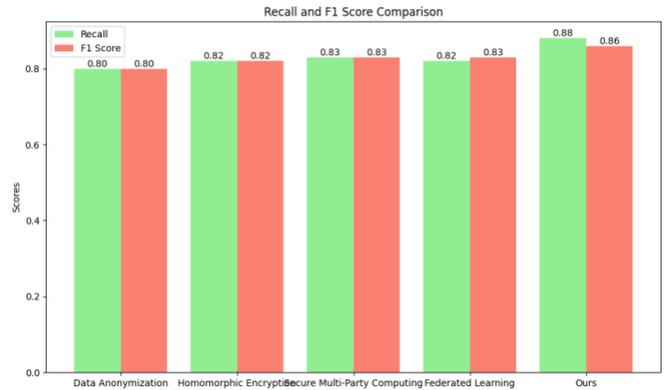

Figure 3 The Recall and F1 Score experimental results

V.  CONCLUSION

The results of this study clearly demonstrate that our proposed algorithm is significantly effective in privacy protection. By applying differential privacy technology, our model successfully protects user data while ensuring the reliability of data analysis results. This method effectively prevents the leakage of sensitive information and maximizes the protection of users' personal privacy when using natural language processing technology. Compared with other privacy protection models, especially data anonymization, which may lead to performance degradation when protecting privacy, our algorithm can maintain good data processing effects while protecting privacy. This achievement provides new ideas for privacy protection technology in data-driven applications and highlights the importance of differential privacy in today's information age. Furthermore, our model not only improves the effectiveness of privacy protection, but also emphasizes the necessity of privacy protection when performing data processing and analysis. In the context of the rapid development of information technology, the security of user data has received increasing attention. Our findings highlight the urgency and importance of privacy protection measures, especially when dealing with text data involving sensitive personal information. By implementing effective privacy protection technologies, businesses and organizations can build user trust, thereby promoting the widespread application and development of technology. In short, our algorithm provides a feasible solution in the field of privacy protection, reflects the importance of achieving a balance between technological progress and user privacy, and indicates that there will be more innovation and in-depth research in this field in the future.

## References

[1] Sousa S, Kern R, "How to keep text private? A systematic review of deep learning methods for privacy-preserving natural language processing," Artificial Intelligence Review, vol. 56, no. 2, pp. 1427-1492, 2023.

[2] Dang, B., Ma, D., Li, S., Qi, Z. and Zhu, E., "Deep learning-based snore sound analysis for the detection of night-time breathing disorders," Applied and Computational Engineering, vol. 76, pp. 109-114, Jul. 2024. doi: 10.54254/2755-2721/76/20240574.

[3] Gao C, Yu J, "Securerc: a system for privacy-preserving relation classification using secure multi-party computation," Computers & Security, vol. 128, p. 103142, 2023.

[4] Vats A, Liu Z, Su P, et al., "Recovering from privacy-preserving masking with large language models," ICASSP 2024-2024 IEEE International Conference on Acoustics, Speech and Signal Processing (ICASSP), IEEE, pp. 10771-10775, 2024.

[5] Lee K J, Jeong B, Kim S, et al., "General Commerce Intelligence: Glocally Federated NLP-Based Engine for Privacy-Preserving and Sustainable Personalized Services of Multi-Merchants," Proceedings of the AAAI Conference on Artificial Intelligence, vol. 38, no. 21, pp. 22752-22760, 2024.

[6] Behera C K, Lakshmi D, Kondurkar I, Enhancing User Privacy in Natural Language Processing (NLP) Systems: Techniques and Frameworks for Privacy-Preserving Solutions, IGI Global, pp. 159-185, 2024.

[7] Peris C, Dupuy C, Majmudar J, et al., "Privacy in the time of language models," Proceedings of the sixteenth ACM international conference on web search and data mining, pp. 1291-1292, 2023.

[8] Akimoto Y, Fukuchi K, Akimoto Y, et al., "Privformer: Privacy-preserving transformer with mpc," 2023 IEEE 8th European Symposium on Security and Privacy (EuroS&P), IEEE, pp. 392-410, 2023.

[9] Petit-Jean T, Gerardin C, Berthelot E, et al., "Collaborative and privacy-preserving workflows on a clinical data warehouse: an example developing natural language processing pipelines to detect medical conditions," medRxiv, pp. 2023.09.11.23295069, 2023.

[10] Edemacu K, Wu X, "Privacy preserving prompt engineering: A survey," arXiv preprint arXiv:2404.06001, 2024.

[11] H. Qin, H. Zheng, B. Wang, Z. Wu, B. Liu, and Y. Yang, "Reducing Bias in Deep Learning Optimization: The RSGDM Approach," arXiv preprint, arXiv:2409.15314, 2024.

[12] H. Zheng, B. Wang, M. Xiao, H. Qin, Z. Wu, and L. Tan, "Adaptive Friction in Deep Learning: Enhancing Optimizers with Sigmoid and Tanh Function," arXiv preprint, arXiv:2408.11839, 2024.

[13] Q. Wang, Z. Gao, T. Mei, X. Cheng, W. Gu, and H. Xia, "Deep Learning-based Multimodal Fusion for Improved Object Recognition Accuracy," in Proc. of the 2024 3rd Int. Conf. on Robotics, Artificial Intelligence and Intelligent Control (RAIIC), pp. 471-474, July 2024.

[14] Z. Zhang, J. Chen, W. Shi, L. Yi, C. Wang, and Q. Yu, "Contrastive Learning for Knowledge-Based Question Generation in Large Language Models," arXiv preprint, arXiv:2409.13994, 2024.

[15] Y. Liang, Y. Zhang, Z. Ye, and Z. Chen, "Enhanced Unsupervised Image Registration via Dense U-Net and Channel Attention," Journal of Computer Science and Software Applications, vol. 4, no. 5, pp. 8-15, 2024.

[16] Z. Wu, H. Gong, J. Chen, Y. Zhu, L. Tan, and G. Shi, "A Lightweight GAN-Based Image Fusion Algorithm for Visible and Infrared Images," arXiv preprint, arXiv:2409.15332, 2024.

[17] Yao J, Li C, Sun K, Cai Y, Li H, Ouyang W, and Li H, "Ndc-scene: Boost monocular 3D semantic scene completion in normalized device coordinates space," Proceedings of the 2023 IEEE/CVF International Conference on Computer Vision (ICCV), pp. 9421-9431, Oct. 2023.

[18] D. Ma, S. Li, B. Dang, H. Zang, and X. Dong, "Fostc3net: A lightweight YOLOv5 based on the network structure optimization," Journal of Physics: Conference Series, vol. 2824, no. 1, p. 012004, Aug. 2024. DOI: 10.1088/1742-6596/2824/1/012004.

[19] X. Yan, Y. Jiang, W. Liu, D. Yi, H. Sang, and J. Wei, "Data-Driven Spatiotemporal Feature Representation and Mining in Multidimensional Time Series," arXiv preprint, arXiv:2409.14327, 2024.

[20] E. Gao, H. Yang, D. Sun, H. Xia, Y. Ma, and Y. Zhu, "Text Classification Optimization Algorithm Based on Graph Neural Network," arXiv preprint, arXiv:2408.15257, 2024.

[21] W. Dai, J. Tao, X. Yan, Z. Feng, and J. Chen, "Addressing Unintended Bias in Toxicity Detection: An LSTM and Attention-Based Approach," in Proc. of the 2023 5th Int. Conf. on Artificial Intelligence and Computer Applications (ICAICA), pp. 375-379, 2023.

[22] W. He, R. Bao, Y. Cang, J. Wei, Y. Zhang, and J. Hu, "Axial Attention Transformer Networks: A New Frontier in Breast Cancer Detection," arXiv preprint, arXiv:2409.12347, 2024.

[23] S. Bo, Y. Zhang, J. Huang, S. Liu, Z. Chen, and Z. Li, "Attention Mechanism and Context Modeling System for Text Mining Machine Translation," arXiv preprint, arXiv:2408.04216, 2024.

[24] M. Sui, J. Hu, T. Zhou, Z. Liu, L. Wen, and J. Du, "Deep Learning-Based Channel Squeeze U-Structure for Lung Nodule Detection and Segmentation," arXiv preprint, arXiv:2409.13868, 2024.

[25] X. Yan, W. Wang, M. Xiao, Y. Li, and M. Gao, "Survival Prediction Across Diverse Cancer Types Using Neural Networks," in Proc. of the 2024 7th Int. Conf. on Machine Vision and Applications, pp. 134-138, 2024.

[26] Q. Wang, Z. Gao, M. Sui, T. Mei, X. Cheng, and I. Li, "Enhancing Convolutional Neural Networks with Higher-Order Numerical Difference Methods," arXiv preprint, arXiv:2409.04977, 2024.

[27] Y. Wu, K. Xu, H. Xia, B. Wang, and N. Sang, "Adaptive Feature Interaction Model for Credit Risk Prediction in the Digital Finance Landscape," Journal of Computer Science and Software Applications, vol. 3, no. 1, pp. 31-38, 2023.